# High redshift AGNs from the 1Jy catalogue and the magnification bias

N. Benítez & E. Martínez-González
Instituto de Física de Cantabria, CSIC-Universidad de Cantabria
and Dpto. de Física Moderna, Universidad de Cantabria
Avda. Los Castros s/n, 39005 Santander, Spain

## ABSTRACT

We have found a statistically significant (99.1 %) excess of red ($O-E>2$) galaxies with photographic magnitudes $E<19.5$, $O<21$ taken from the APM Sky Catalogue around $z\sim 1$ radiosources from the 1Jy catalogue. The amplitude, scale and dependence on galaxy colours of the observed overdensity are consistent with its being a result of the magnification bias caused by the weak gravitational lensing of large scale structures at redshift $z\approx 0.2-0.4$ and are hardly explained by other causes, as obscuration by dust.

## 1. Introduction

Several authors have confirmed the existence of statistically significant associations of foreground objects, usually bright galaxies and clusters, with high redshift AGNs (Webster et al. 1988, Fugmann 1990, Hammer & Le Fèvre 1990, Thomas et al. 1994, Benítez et al. 1995, Seitz & Schneider 1995). These associations have been explained as a consequence of the gravitational magnification bias (Narayan & Wallington 1989). This phenomenon arises when background sources (like the high redshift AGNs) whose line of sight crosses a matter positive fluctuation are amplified by a magnification factor $\mu$, what makes them more easy to detect in that patch of the sky. Besides, the surface density of background objects behind the lens is diluted by a factor $\mu^{-1}$. The resulting effect is described by the expression $n_A(S)=n'_A(S/\mu)/\mu$ (Narayan 1989), where $n_A(S)$ is the observed cumulative number density, $\mu$ is the magnification, and $n'_A$ is the unlensed number density which usually has the form $n'_A(S') \propto (S')^\alpha$, $S'$ being the AGN flux in absence of magnification. This yields $n_A(S) \propto \mu^{-\alpha-1} S^\alpha$ and from this expression it is evident that depending on the slope $\alpha$, we can observe an excess ($n_A > n'_A$) or even a deffect ($n_A < n'_A$) of background sources around the lenses. Galaxies and clusters trace the matter fluctuations (up to a bias factor $b$), and this means that depending on the slope of the number counts of background sources, we may observe positive, null or negative statistical associations of these foreground objects with background sources like distant AGNs (see Bartelmann 1995, Seitz & Schneider 1995).

The scales at which the mentioned associations seem to be present vary from a few arcseconds (Webster et al. 1988, Thomas et al. 1994) to $\sim 30'$ (Fugmann 1990). This seems to indicate that the size of the lenses causing the magnification bias varies from galaxies to matter overdensities on scales at least as large as clusters of galaxies or even larger (Pei 1995, Bartelmann 1995). Therefore, the study of this effect provides information about a large range of matter fluctuation scales.

Radio-selected samples of high-redshift objects present a rather steep slope in their number counts-flux distribution. This type of samples is therefore very appropiate for trying to detect excesses of foreground objects due to the magnification bias. The optimal choice would be using a sample of background objects which is flux limited in two or more uncorrelated wavelength bands, what can steepen the number counts–flux relation considerably (Borgeest et al. 1991, Bartelmann 1995). In this work we shall study the distribution of galaxies taken from the APM Sky Catalogue around a sample of AGNs from the 1Jy catalogue (Stickel et al. 1994).

## 2. The Data

The 1Jy catalogue of radio-sources is described in Stickel et al. 1994. It has been often and succesfully used to prove the existence of correlations between several catalogues of foreground objects and distant AGNs (Fugmann 1990, Seitz & Schneider 1995, Bartelmann & Schneider 1994). Our initial sample of AGNs is formed by all the QSOs and radiogalaxies contained in the catalogue with measured redshifts $0.5 < z < 1.5$ and with declination $\delta > 0$, what yields 91 AGNs.

The APM Sky Catalogue is described in Irwin et al. 1994. It provides the user with files containing all the objects found by the detection algorithm PISA on the scans of the Palomar Sky Survey plates. The files contain, among other data, the classification of the detected objects (star/galaxy) and their $O$ and $E$ photographic magnitudes. The $O$ band is equivalent to $U+B$ and the $E$ band corresponds to a narrow $R$ (Irwin & McMahon 1992). We asked for the files containing all the objects within a $90' \times 90'$ box centered on each AGN of the sample. Several fields (8) were not available at the APM Sky Catalogue database because of the low galactic latitude of the plates containing them ($|b| < 20°$). Some of the $90' \times 90'$ fields extend over different plates; besides, the plates have $10'$ overlaps, so some AGNs can be found on two or three different plates. In order to make the fields as homogeneus as possible, each of our fields contains data from one sole plate, namely from the one which contains the biggest portion of that field. Using only one plate for each field means therefore that those fields which happened to be close to a plate border are incomplete. In order to have enough statistics to calculate accurately the average density within each field, we discarded several fields (10) because they did not have at least 2/3 of the required $5400'' \times 5400''$ surface within one plate.

We finally have 73 fields which contain all the objects classified as galaxies in the APM Sky Catalogue in the red or blue band and with $O < 19.5$ and $E < 21$. The colours are consistent to $\pm 0.2$ magnitudes and the magnitude faint end zero-points are consistent to $\pm 0.25$ magnitudes (Irwin et al. 1994). Our chosen magnitude limits are 0.5 mag brighter than the limiting magnitudes on each band quoted by Irwin and collaborators. Number counts-magnitude plots show that they can be reasonably considered as a sort of completion limits in the sense that fainter than these limits the number counts-magnitude function suddenly gets steeper and the spurious detections begin to completely dominate the number counts. In order to exclude as many spurious galaxies as possible, we require our objects to be detected and classified as galaxies in both the red *and* blue bands. This discards $\approx 40 - 50\%$ of the objects at magnitudes below our completion limits. At brighter magnitudes ($E < 18$), most of them are objects classified as galaxies in one band and as stars or merged objects in the other. This is not surprising if we remember that the classification of the objects in the APM Sky Catalogue is an automated procedure and that the pixel resolution is rather low. At fainter magnitudes we are also losing many galaxies that due to their colours cannot be detected in both bands. Nevertheless, the followed procedure is necessary in order to obtain an enough homogeneus sample. Our final galaxy sample is therefore formed by the fields centered on 73 AGNs and contains $\approx 35,000$ objects classified as galaxies in the APM Sky Catalogue in the red and blue band and with $E < 19.5$ and $O < 21$.

To test for the existence of systematic gradients on the fields, we divided each field into four $45' \times 45'$ boxes and added the number of galaxies found on each of these boxes over all the fields, after correcting by the 'useful' surface of each box on each field. The differences among the number of objects found in the four boxes agrees well with a $\sqrt{N}$ poissonian dispersion. Another significant test is to consider the number of objects within the central part of all the fields containing half of the total $5400'' \times 5400''$ surface (excluding a central $1000'' \times 1000''$ region which contains most of the excess) and compare it with the number of objects on the outer half, after correcting by the useful surface. The difference is again consistent with the poissonian $\sqrt{N}$. These two tests are a strong confirmation of the homogeneity of the fields on average.

## 3. Main results and discussion

In order to look for density enhancements on a certain scale $l$, we shall use the following method. We divide the fields into boxes of size $l^2$ and define the density enhancement on each box as

$$q_{ij} = \frac{\sum_f n_{ij}^f}{\sum_f \frac{s_{ij}^f N^f}{S^f}}$$

where $n_{ij}^f$ and $s_{ij}^f$ are the number of galaxies and the useful surface respectively in the box $ij$ of the field $f$, $N^f$ is the total number of galaxies of a field and $S^f$ its total useful surface.

We estimate the statistical significance of the result in a straightforward and distribution–independent way through the empirical probability $p$, $p = N_{less}/N_{total}$ where $N_{less}$ is the number of boxes (not including the central one) with a value of $q$ less or equal than that of the central box and $N_{total}$ is the total number of boxes in the field. This estimator is quite conservative since it tends to underestimate the real signification of the excess as the maximal signification is always $p_{max} = (N_{total} - 1)/N_{total}$ no matter how far the value of the central box is from the average. The empirical distribution of $q$ values is asymmetric with a tail toward high values of $q$. We find that $p$ is always similar or smaller than the value of the significance obtained using the empirical rms (which is a factor $1.2 - 1.4$ the one deduced from poissonian statistics, Benítez et al. 1995) and assuming a gaussian distribution. In any case we think that our galaxy fields are large enough to provide meaningful values of $p_{max}$ for the considered angular scales. The main strenght of this method is thus its simplicity and the fact that it does not make any assumptions on the probability distribution of the variable q.

The results for scales of 500″and 1000″and different colours are presented in table 1. It can be seen that if we consider *all* the galaxies with $E < 19.5$ and $O < 21$ we do not find a very significant excess. However, the value of the density enhancement strongly increases for red galaxies with $O - E > 2$ and reaches a value of $q = 1.298 \pm 0.114$ (rms error), the highest of all the $500'' \times 500''$ boxes in the field, corresponding to an empirical $p$ of 99.1%. This behaviour agrees well with elementary physical considerations. Background objects with redshift $z_s \approx 0.5 - 1.5$ are most efficiently magnified by foreground objects with $z_l \approx 0.2 - 0.4$ (Schneider et al. 1992). Most types of $M^*$ galaxies at these redshifts will have colours $O-E \gtrsim 2$ ($H_0 = 50, \Lambda = 0, \Omega = 1$. $M^*$ and k-corrections are taken from Driver et al. 1994, and we have used the relationship $O - E \approx 2(B-V)$ quoted in Irwin & McMahon 1992 ). Therefore, selecting galaxies with $O-E > 2$ means isolating the ones most likely to be tracing the potential wells acting as lenses for the AGN of our sample, what evidently maximizes the excess, otherwise diluted among the objects at lower redshifts which belong to structures not contributing so strongly to the lensing effect. The scale at which the density enhancement is more significant, $500''$, corresponds to $\approx 2-3$ Mpc at redshifts of $z = 0.2 - 0.4$.

In order to show the dependence of the excess with the angular scale we plot the cross-correlation AGN-galaxy

$$\omega_{AG}(\theta) = \frac{\sum_f N_G^f(\theta)}{\sum_f N_{exp}^f(\theta)} - 1$$

where $N_G^f(\theta)$ is the number of galaxies found on the annular bin $\theta$ of the field $f$ and $N_{exp}^f(\theta)$ is its expectation based on the number of objects on the whole field normalized to the useful surface of the bin, for the galaxies with $O < 21, E < 19.5$ and $O-E > 2$ (figure 1) as a function of the radius in arcsec. The error bars are poissonian $\sqrt{N}$.

The previous results should be confronted with the theoretical expectations of galaxy-QSO cross-correlations $\omega_{QG}(\theta)$ produced by weak and moderate gravitational lensing. Recently Bartelmann (1995) has calculated the expected cross-correlation for samples of galaxies with different depths and AGN catalogues simulating the 1Jy, within a CDM cosmogony. The value of the parameter q on a scale of 500 $''$ ($q-1$ on a scale $l$ coincides with the expression for the cross-correlation amplitude within a radius $l/2$, except that the former is defined on a square box and the latter on a circular surface) in the case of no colour constraint on the APM galaxies (see table1) agrees marginally with the results of Bartelmann 1995. In any case, due to the small number of AGNs considered and the rather bright magnitude limit of the APM galaxies the amplitude is not significantly different from zero as expected by Bartelmann. However it is noteworthy that, as shown above, considering only the reddest galaxies produces a strong enhancement of $\omega_{AG}$ as we select the objects tracing the potential wells which more effectively act as lenses for the 1Jy sample.

We have also tried to measure the effect of imposing a double flux limit by setting a cutoff on the optical magnitude of the AGNs. Although it seems to be a tendency towards greater excesses for AGNs with brighter optical magnitudes, this constraint reduces the sample and lowers the statistical significance, what makes the found results not conclusive. Larger samples are needed in order to better establish this effect.

The observed overdensity of foreground galaxies around high-redshift AGN seems to be a result of the gravitational lensing of the large scale structure and is difficult to explain by other causes, as obscuration by dust. In fact, it is easy to see that this last effect should produce the opposite of what is observed: the objects forming the excess around high redshift AGNs would tend to be slightly bluer than the average.

This research made use of the APM measurements of the POSS plates, accesible on line from IoA at Cambridge. The 1Jy catalogue was obtained from the Astronomer's Bazaar, the CDS service for astronomical catalogues accesible on line from Strasbourg. We thank J.L. Sanz and J. Silk for interesting comments and discussions and E. Stengler-Larrea for technical support. NB and EMG acknowledge financial support from the Spanish DGICYT, project PB92-0741. NB acknowledges a Spanish M.E.C. Ph.D. scholarship.

## REFERENCES

Bartelmann, M., 1995, A&A, in press

Borgeest, U., von Linde, J., Refsdal, S., 1991, A&A, 251, L35

Bartelmann, M. & Schneider, P. 1994, A&A, 284, 1

Benítez, N., Martínez-González, E., González-Serrano & Cayón L. 1995 AJ, 109, 935

Driver, S.P., Phillipps, Davies, J.I., Morgan, I. & Disney, M.J., 1994 MNRAS, 266, 155

Fugmann, W., 1990, A&A, 240, 11

Hammer F., Le Févre O., 1990, ApJ, 357, 38

Irwin, M., & McMahon, R. 1992, Gemini, 37

Irwin, M., Maddox, S. & McMahon, R. 1994, Spectrum, Newsletter of The Royal Observatories, 2, 14

Narayan, R., 1989, ApJ, 341,L1

Narayan, R., Wallington, S. 1993, in: Gravitational Lenses in the Universe, Proc. 31st Liège Astroph. Coll., eds. J. Surdej et al.

Pei, Y., 1995, ApJ, 440, 485

Schneider, P., Ehlers, J., Falco, E.E., 1992, Gravitational Lenses (Heidelberg: Springer)

Stickel, M., Meisenheimer K.,& Kühr, H. 1994, A&AS, 105, 211

Seitz, S. & Schneider, P. 1995, A&A, in press

Thomas, P.A., Webster, R.L., & Drinkwater, M.J. 1994, to appear in MNRAS

Wu, X.P, 1994, A&A, 286, 748

Webster, R.L., Hewett, P.C., Harding, M.E., Wegner, G.A., Nature, 336, 358

This preprint was prepared with the AAS LaTeX macros v3.0.

FIGURE CAPTIONS

Figure 1. The cross-correlation function $\omega_{AG}$ between 1Jy AGNs with $0.5 < z < 1.5$ and APM galaxies with $E < 19.5$, $O < 21$ and $O - E > 2$. Error bars are $1\sigma$ $\sqrt{N(\theta)}$.

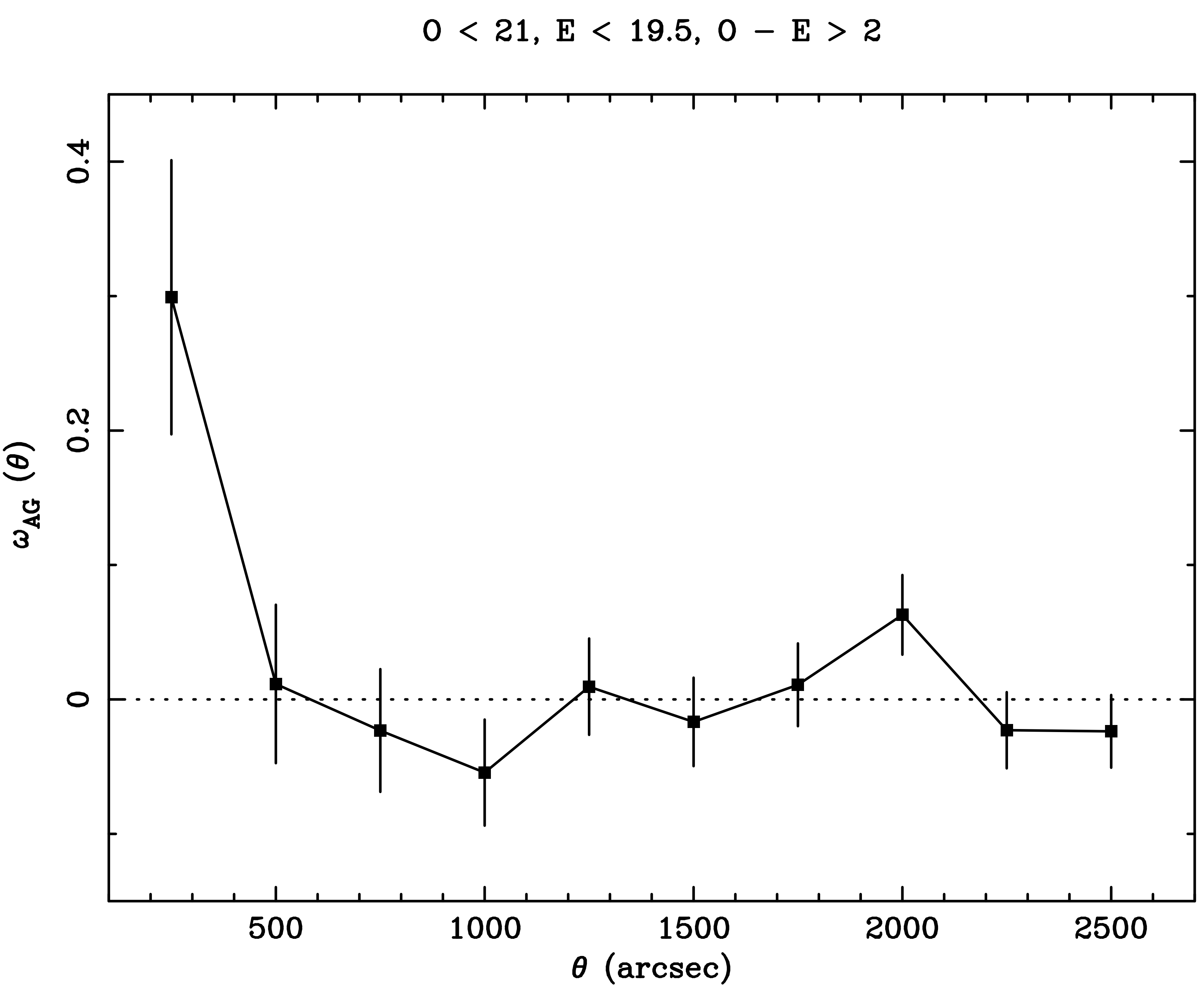
O < 21, E < 19.5, O − E > 2
0.4
0.2
0
$\omega_{AG}$ ($\theta$)
500
1000
1500
2000
2500
$\theta$ (arcsec)

TABLE 1

Data for the central box

| $O-E$ | 500 arcsec | 1000 arcsec |
|---|---|---|
| all | $1.109 \pm 0.078$, 91.4% | $1.028 \pm 0.041$, 71.7% |
| $> 1$ | $1.122 \pm 0.082$, 92.3% | $1.039 \pm 0.047$, 83.9% |
| $> 2$ | $1.298 \pm 0.114$, 99.1% | $1.089 \pm 0.059$, 91.7% |

Note.— Values of $q$ and $p$ for different scales and colours. The errors in $q$ are the empirical rms over all the boxes in the field